\documentclass [] {aa} 
\usepackage{epsfig, graphicx}
\usepackage{natbib}

\bibpunct{(}{)}{;}{a}{}{,}

\begin{document}

%\thesaurus{06(08.01.1; 08.03.02; 08.16.2)}

\title{The CORALIE survey for southern extra-solar planets\thanks{Based on
    observations collected at the La Silla Observatory, ESO (Chile),
    with the {\footnotesize CORALIE} spectrograph at the 1.2-m
    Euler Swiss telescope and with the Str{\"o}mgren
    Automatic Telescope (SAT).}}
\subtitle{IX. A 1.3-day period brown dwarf disguised as a planet}

\author{
      N.C.~Santos \inst{1}
 \and M.~Mayor \inst{1}
 \and D.~Naef \inst{1}
 \and F.~Pepe \inst{1}
 \and D.~Queloz \inst{1}
 \and S.~Udry \inst{1}
 \and M.~Burnet \inst{1}
 \and J.V.~Clausen \inst{2} 
 \and B.E.~Helt \inst{2} 
 \and E.H.~Olsen \inst{2}
 \and J.D.~Pritchard \inst{3}
}

\offprints{Nuno C. Santos, \email{Nuno.Santos@obs.unige.ch}}

\institute{
	Observatoire de Gen\`eve, 51 ch.  des 
	Maillettes, CH--1290 Sauverny, Switzerland
     \and
	Niels Bohr Institute for Astronomy, Physics, and 
	Geophysics; Astronomical Observatory, Juliane Maries 
	Vej 30, DK-2100 Copenhagen {\O}, Denmark
     \and
	European Southern Observatory, Santiago, Chile
 }

\date{Received / Accepted } 

\titlerunning{A brown dwarf around HD\,41004\,B} 

%--------------------------------------------------------------------------

\abstract{
{ In this article we present the case of HD\,41004\,AB, a system composed of a K0V star
and a 3.7-magnitude fainter M-dwarf companion. 
We have obtained 86 CORALIE spectra of this system with the goal of obtaining precise
radial-velocity measurements. Since HD\,41004\,A and B are separated by only 0.5\arcsec, in every 
spectrum taken for the radial-velocity measurement, we are observing the blended spectra 
of the two stars. An analysis of the measurements has revealed a velocity variation 
with an amplitude of about 50\,m\,s$^{-1}$ and a periodicity of 1.3\,days. 
This radial-velocity signal is consistent with the 
expected variation induced by the presence of a companion to either HD\,41004\,A or HD\,41004\,B,
or to some other effect due to e.g. activity related phenomena.
In particular, such a small velocity amplitude could be the signature of the
presence of a very low mass
giant planetary companion to HD\,41004\,A, whose light dominates the spectra.
The radial-velocity measurements were then complemented with a photometric campaign and
with the analysis of the bisector of the CORALIE Cross-Correlation Function (CCF).
While the former revealed no significant variations within the observational precision of 
$\sim$0.003-0.004\,mag (except for an observed flare event), the bisector analysis
showed that the line profiles are varying in phase with the radial-velocity.
This latter result, complemented with a series of simulations, has shown that we 
can explain the observations by considering that HD\,41004\,B has 
a brown-dwarf companion orbiting with the observed 1.3-day period. 
As the spectrum of the fainter HD\,41004\,B ``moves'' relative to the one of HD\,41004\,A (with 
an amplitude of a few km\,s$^{-1}$),
the relative position of the spectral lines of the two spectra changes, thus changing the blended 
line-profiles. This variation is large enough to explain the observed radial-velocity and
bisector variations, and is compatible with the absence of any photometric signal. If confirmed, 
this detection represents the first discovery of a
brown dwarf in a very short period (1.3-day) orbit around an M dwarf.
Finally, this case should be taken as a serious warning about the importance of
analyzing the bisector when looking for planets using radial-velocity techniques.}
\keywords{techniques: radial velocities -- 
binaries: visual -- 
binaries: spectroscopic -- 
stars: brown dwarfs -- 
stars: exoplanets -- 
stars: individual: HD\,41004}
}

\maketitle

\section{Introduction}

{ Radial-velocity techniques have so far unveiled about 80 planetary companions around} 
solar type dwarfs\footnote{See e.g. obswww.unige.ch/$\sim$udry/planet/planet.html}. 
The most precise instruments currently available for planet searches
can measure the velocity of a star in the direction of the line-of-sight
with a precision of the order of 2-3\,m\,s$^{-1}$ \citep[e.g.][]{Que01a,But01,Pep02}, but even
higher precision is expected from instruments available in the near future
\citep[e.g. HARPS -- ][]{Pep00}. { This will definitely allow the discovery of 
lower mass and longer period planets}, that remained
undetected up to now due to the low amplitude of
the induced radial-velocity variation.

The gain in precision will, however, bring to light some of the
limitations of the radial-velocity method. It is well known, for example,
that radial-velocity ``jitter'' with amplitudes up to a few tens of m\,s$^{-1}$ is 
expected to result from the presence of strong photospheric features like spots
or convective inhomogeneities,
associated with chromospheric activity phenomena \citep[][]{Saa97,Saa98,San00}. 
The presence of spots can even induce a periodic radial-velocity
signal similar to the one expected from the presence of a planet. 
This is the case for \object{HD\,166435} \citep[][]{Que01b}, a star presenting a 
radial-velocity signal with a period of about 3.8-days, but showing both photometric and bisector
variations with the same periodicity.

\begin{table}
\caption[]{
Stellar parameters for \object{HD\,41004\,A}}
\begin{tabular}{lcc}
\hline
\hline
\noalign{\smallskip}
Parameter  & Value & Reference \\
\hline \\
$Spectral~type$  & K1V/K2V & Hipparcos \citep[][]{ESA97}/  \\
                 &         & {\it uvby} (see Sect\,\ref{sec:photom})  \\
$Parallax$~[mas]  & 23.24 $\pm$ 1.02 & Hipparcos \citep[][]{ESA97}\\
$Distance$~[pc]  & 43 & Hipparcos \citep[][]{ESA97} \\
$m_v$  & 8.65 & Hipparcos \citep[][]{ESA97}\\
$B-V$  & 0.887 & Hipparcos \citep[][]{ESA97} \\
$T_\mathrm{eff}$~[K]  & 5010 & See text \\
$\log{g}$~[cgs]  & 4.42 & See text \\
$M_\mathrm{v}$  & 5.48 & -- \\
$Luminosity~[L_{\sun}]$ & 0.65 & \citet{Flo96} \\
$Mass~[M_{\sun}]$  & $\sim$0.7 & -- \\[5.0pt]
$\log{R'_\mathrm{HK}}$  & $-$4.66 & \citet[][]{Hen96} \\
$Age$~[Gyr] & 1.6 & \citet[][]{Don93} \\
$P_{\mathrm{rot}}$~[days]   & $\sim$27 &  \citet[][]{Noy84} \\[5.0pt]
$v\,\sin{i}$~[km\,s$^{-1}$] & 1.22 & CORALIE \\
$\mathrm{[Fe/H]}$  & $-$0.09/$+$0.10 & {\it uvby}/CORALIE  \\
\hline
\noalign{\smallskip}
\end{tabular}
\label{tab1}
\end{table}

The case of \object{HD\,166435} illustrates very well the need to
confirm, at least for the shortest period cases, that the radial-velocity signature is
indeed due to the presence of a low mass companion, and not due to some kind of intrinsic phenomena. As shown by \citet[][]{Que01b},
the use of photometric data and bisector analysis was crucial to 
clarify the origin of the radial-velocity variations observed on this star. 

In this paper we present the case of \object{HD\,41004},
a visual double system consisting of a K1V-M2V pair (A and B components). This system
was found to present a radial-velocity signature similar to the one expected as if
the K1 dwarf had a very low mass planetary companion in a 1.3-day
period orbit. Although the photometric data revealed no significant
photometric variations, an analysis of the Bisector Inverse Slope (BIS) of the
CORALIE Cross-Correlation Function (CCF) \citep[][]{Que01b} revealed a periodic variation in
phase with the radial-velocity signal. In the following sections we will show that the 
radial-velocity variation is in fact not a result of the periodic motion of the A component, 
but of the Doppler motion of the spectrum of the B component due to the presence of
a brown-dwarf companion. The results strongly caution about the need to use
methods capable of detecting line asymmetries, like the bisector analysis, when dealing 
with high-precision planet searches with radial-velocity techniques.

\section{The case of \object{HD\,41004}}

\subsection{Stellar characteristics}

\begin{figure}[t]
\psfig{width=\hsize,file=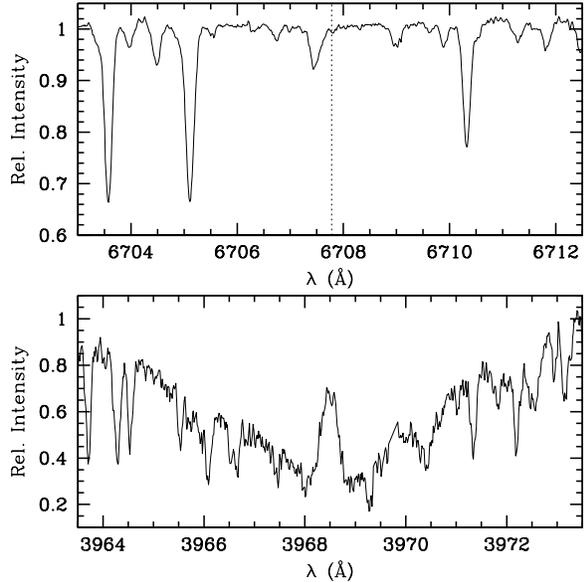}
\caption[]{
{\it Upper panel}: Added CORALIE spectrum in the Lithium line region for \object{HD\,41004}. 
The plot reveals no clear Li detection for this star. 
{\it Lower panel}: Added CORALIE spectrum in the \ion{Ca}{ii} H line central region for 
\object{HD\,41004}. The strong emission near line center suggests that the star is 
chromospherically active}
\label{fig1}
\end{figure}

In the Hipparcos catalogue \object{HD\,41004} (\object{HIP\,28393}, \object{CD\,$-$48\,2083}) 
is presented as the A component of a double system, where the 3.68\,mag fainter 
companion is at a separation of about 0.5\arcsec. From the magnitude difference,
and considering that both stars are at the same distance (i.e. that they constitute
a physical double system) we can deduce that \object{HD\,41004\,B} is probably a M2 dwarf.
In the rest of this paper we will denote by
\object{HD\,41004\,AB} the system composed by \object{HD\,41004\,A} and \object{HD\,41004\,B}. 

{ 
The small angular separation between the two components implies that
we cannot separate (with the instrumentation used) the light coming from the two
stars. However, given the relatively ``large'' magnitude difference, we do not expect
the flux coming from \object{HD\,41004\,B} (about 3\% of that coming from the A component) 
to affect significantly the determination of the stellar parameters of 
\object{HD\,41004\,A}. All the parameters described below are thus, 
in a good approximation, those of \object{HD\,41004\,A} (the star whose 
light dominates).
}

At a distance of $\sim$43~parsec -- $\pi$=23.24$\pm$1.02\,mas -- 
\citet[][]{ESA97}, \object{HD\,41004\,A} 
is a K1 dwarf shining with a visual magnitude V=8.65 in the southern constellation Pictoris 
(the Painter's Easel). Its colour index is $B-V\,=\,0.887$, as listed by the Hipparcos catalogue
and its absolute magnitude $M_{\mathrm{v}}$=5.48. The basic stellar parameters are 
summarized in Table\,\ref{tab1}.

There are not many references in the literature on \object{HD\,41004\,A}, and
no precise spectroscopic analysis is available for this star.
Using the $b-y$ colour from \citet[][]{Ols94b}\footnote{A similar $b-y$ was obtained by 
us -- see Sect.\,\ref{sec:photom}.} and the calibration of
\citet[][]{Alo96} we obtain a temperature of 5030\,K, 100\,K higher than the
4930\,K obtained from the same colour but using the calibration of \citet[][]{Ols84}.
Using the $B-V$ colour we would obtain T$_{\mathrm{eff}}$=5075\,K from the calibration of
\citet[][]{Flo96}. We have decided to use an average value in the following of this paper 
(T$_{\mathrm{eff}}$=5010\,K). Photometric calibrations can also be used to 
estimate the surface gravity of a dwarf.
Using the calibration of \citet[][]{Ols84} we have obtained
$\log{g}$=4.54, slightly higher than (but compatible with) the value
of 4.30 obtained from the Hipparcos parallax \citep[e.g.][]{All99} -- again, an average 
value was considered. The obtained T$_{\mathrm{eff}}$ and $\log{g}$ are compatible with a
spectral classification of K1V \citep[][]{ESA97}.

From the photometric calibration of \citet[][]{Sch89} we have obtained a value of 
[Fe/H]=$-$0.09. Another estimation of the metallicity can be obtained from the
analysis of the Cross-Correlation Function (CCF) of CORALIE. In fact, its equivalent width (or surface) is very well correlated with the metallicity of the star and its colour $B-V$
(Sect.\,\ref{apendix2}). Using this method we have derived a higher 
value of [Fe/H]=$+$0.10. 

Using a metallicity of $+$0.10, the temperature of 5010\,K, and its absolute magnitude,
we can estimate a mass of $\sim$0.7\,M$_{\sun}$ for \object{HD\,41004\,A} using the
isochrones of \citet[][]{Sch92}. 

\begin{table}
\caption[]{Elements of the fitted orbit }
\begin{tabular}{lrll}
\hline
\hline
\noalign{\smallskip}
$P$             & 1.3298                & $\pm$    0.0002                 & d\\
$a_1\,\sin i$   & 0.0009                & $\pm$    0.0000                 & Gm\\
$T$             & 2452200.43            & $\pm$    0.13                   & d\\
$e^\dagger$     & 0.07                  & $\pm$    0.04                   &  \\
$V_r$           & 42.532                & $\pm$    0.002                  & km\,s$^{-1}$\\
$\omega^\dagger$ & 23                   & $\pm$    35                     & degr \\ 
$K_1$           & 50                    & $\pm$    2                      & m\,s$^{-1}$ \\
$f_1(m)$        & $0.1741\cdot10^{-10}$ & $\pm$    $0.0224\cdot10^{-10}$  & M$_{\sun}$\\ 
$\sigma(O-C)$   & 14                    &                                 & m\,s$^{-1}$  \\    
$N$             & 86                    &                                 &  \\
\noalign{\smallskip}
\hline
\end{tabular}
\\ $^\dagger$ Consistent with a circular orbit according to the \citet{Luc71} test.
\label{tab2}
\end{table}

The CCF can also be used to determine the projected rotational velocity $v\,\sin{i}$ 
of a star. From the calibration presented in Sect.\,\ref{apendix1}, and 
taking the average Gaussian width of the CCF for our
star (4.36\,km\,s$^{-1}$), we have obtained a value of $v\,\sin{i}$=1.22\,km\,s$^{-1}$,
similar to the 1.54\,km\,s$^{-1}$ obtained from the CORAVEL CCF using the 
calibration of \citet[][]{Ben84}.

\begin{figure}[t]
\psfig{width=\hsize,file=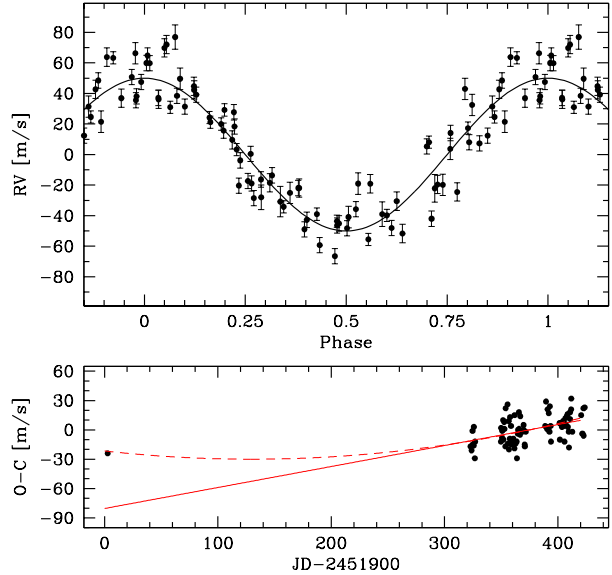}
\caption[]{
{\it Upper panel}: Phased CORALIE radial-velocity measurements (subtracted by the
$\gamma$-velocity) for \object{HD\,41004\,AB} and best Keplerian fit. 
{\it Lower panel}: $O-C$ residuals of the CORALIE radial-velocity data over time. The lines
represent the best linear fit without considering the first measurement (solid line) and
a parabolic fit (dashed line). The linear fit has a
slope of 0.2\,m\,s$^{-1}$\,day$^{-1}$}
\label{fig2}
\end{figure}

An analysis of the \ion{Ca}{ii} H line core of \object{HD\,41004\,A} reveals a strong emission,
suggesting that the star is chromospherically active (see Fig.\,\ref{fig1}, lower panel). This is
confirmed by the value of $\log{R'_\mathrm{HK}}$=$-$4.66 obtained by \citet[][]{Hen96},
and by the fact that a flare event was observed in photometry (see Sect.\,\ref{sec:photom}).
From this value, and using the calibration of \citet[][]{Noy84}, we can estimate the
rotational period to be $\sim$27 days, a value perfectly compatible with the low $v\,\sin{i}$ 
measured. Using the calibration of \citet[][]{Don93} 
\citep[also presented in][]{Hen96}, based on the chromospheric activity level 
we derive an age of 1.6\,Gyr for this star. 

In Fig.\,\ref{fig1} (upper panel) we show the 6708\,\AA\ Li-line region for \object{HD\,41004\,AB}.
There is some hint of the presence of Li, but the small feature at the position of the Li line
has an equivalent width of less than 2\,m\AA. Considering the stellar parameters
of our star, this implies an upper limit for the Li abundance of $\log{N(Li)}$$<$0.0 (this was
derived from a LTE spectral analysis using the line abundance code MOOG \citep{Sne73}, 
and a grid of Kurucz \citep{Kur93} ATLAS9 atmospheres).
Given the low temperature of our star, this limit cannot be used as a strong 
age constraint; the age derived from the activity level is nevertheless compatible with
such a low value for the Li abundance \citep[e.g.][]{Jon99}.

\subsection{Radial-velocity data}
\label{sec:rv}

\object{HD\,41004\,AB} is included in the sample observed in the context of the Geneva
extra-solar planet search programme, at La Silla Observatory (Chile), using the CORALIE 
spectrograph \citep[][]{Udr00}. From December 2000 to February 
2002 we have obtained a total of 86 precise radial-velocity measurements\footnote{Available 
in electronic form at the CDS.} of this system. The average individual 
photon-noise precision of the radial-velocity measurements is 
5.5\,m\,s$^{-1}$.

{
Since the angular separation of the two stars
is much smaller than the diameter of the CORALIE fiber, in each spectrum
obtained we are seeing the blended spectra of \object{HD\,41004\,A} and \object{HD\,41004\,B}.
This means that the resulting radial-velocity is not the one
of any of the two individual stars, but rather it corresponds to a weighted average of the
radial velocities of the two components (although it will be closer to the
radial velocity of the A component, since the M dwarf is ``only'' contributing to
$\sim$3\% of the light). 
}

The analysis of the radial-velocity revealed a variation with a period of $\sim$1.3-days,
and an amplitude of 50\,m\,s$^{-1}$ (see Table\,\ref{tab2}). 
{ A fit to the data (Fig.\,\ref{fig2} -- upper panel) suggested that the 
radial-velocity variation could be due to the presence of a low mass (0.25\,M$_{\mathrm{Jup}}$) 
planetary companion orbiting \object{HD\,41004\,A} at a distance of only about 
0.02\,A.U. (i.e. $\sim$5.5 stellar radius, considering that \object{HD\,41004\,A} has a radii 
typical for a K1 dwarf -- 0.8\,R$_{\sun}$) in a circular trajectory.}
Although this possibility was very tempting, as we will see below the planetary explanation 
is not the best. 

The relatively high $O-C$ residuals of the fit (14\,m\,s$^{-1}$)
are quite difficult to explain in the light of activity phenomena, since we do not
expect them to induce such a high radial-velocity ``jitter'' in
a slow rotator K dwarf \citep[][]{Saa98,San00}. Actually, part of the noise seems to be coming 
from the existence of some radial-velocity drift. An analysis of the $(O-C)$ 
residuals to the fit reveals a clear trend (considering the last group of 
points only), with a gradient 
of 75\,m\,s$^{-1}$\,yr$^{-1}$ -- (Fig.\,\ref{fig2} -- lower panel).
We note that a comparison with old CORAVEL measurements, as well as with the first (isolated) 
CORALIE measurement, does not corroborate this trend, i.e. the old measurements 
(see Fig.\,\ref{fig2}, lower panel) are not aligned with a linear fit to the residuals of the
last group of points. This result is compatible with the 
fact that the radial-velocity signal passed through a minimum. 
{ Finally, some noise might also be coming from the influence of the companion
to \object{HD\,41004\,A}. As shown by \citet[][]{Pep00b}, the presence of a close
companion might introduce some noise in the computed radial-velocity due to 
variations of the flux ratio of the two stars within the CORALIE fiber
as a function of the seeing conditions.
This noise depends on the separation between 
the two objects, their magnitude difference, and their radial-velocity difference. 
For this case we do not expect, however, a strong effect, mostly because of the small angular 
separation between the two objects (the angular separation of the two stars is much smaller than
the diameter of the CORALIE fiber, and thus the influence should be always similar 
and fairly independent e.g. of the seeing conditions).}

As mentioned above, \object{HD\,41004\,B} (a M2 dwarf) is located at an angular 
separation of $\sim$0.5\arcsec. At the distance of \object{HD\,41004\,AB} this corresponds to 
a projected separation of $\sim$21\,A.U., considering that the two stars form a real system\footnote{This is 
corroborated by the simulation presented below, that implies that both stars have similar
radial-velocities.}. This value corresponds to the minimum ``real separation'' (since
we are not able to know the distance between the two stars in the direction of the line-of-sight)
and implies a period of $\sim$90\,yr, using a mass of 0.7\,M$_{\sun}$ and 0.4\,M$_{\sun}$ 
for \object{HD\,41004\,A} and B, respectively. Considering a circular orbit, the
amplitude of the variation has a value of 2.5\,km\,s$^{-1}$, and the
maximum derivative of the velocity would be 170\,m\,s$^{-1}$\,yr$^{-1}$. This value 
is perfectly compatible with the observations, suggesting that the radial-velocity trend
observed has its origin in the motion of \object{HD\,41004\,A} around its lower mass companion \object{HD\,41004\,B}. But we cannot exclude that (at least part of) the observed trend might 
have a different origin, like e.g. the presence of a low mass 
companion to \object{HD\,41004\,A}.

\begin{figure}[t]
\psfig{width=\hsize,file=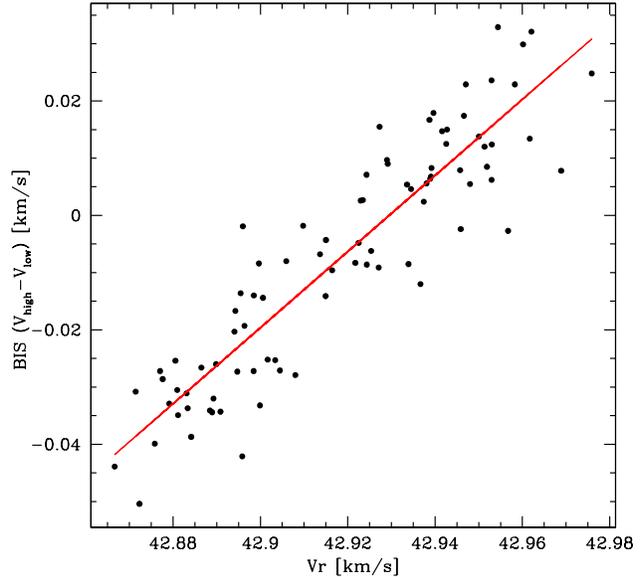}
\caption[]{Measured radial-velocity vs. BIS for HD\,41004. The plot shows the
two variables vary in phase. The best linear fit is shown.
The slope of the line has a value of 0.67}
\label{fig3}
\end{figure}

\subsection{Bisector analysis}
\label{sec:bisector}

With such a short period ``orbit'' (and further knowing that the star is 
chromospherically active), it was extremely important to analyze the bisector 
of the CCF for this object. Such an
analysis, as presented in \citet[][]{Que01b},
is in fact an excellent way of discriminating between radial-velocity variations due to
changes in the spectral line shapes from real variations due to the orbital motion of
the star. In a few words, we compute the (bisector) velocity for 10 different levels of
the Cross-Correlation Function\footnote{We can define 12 levels in total, i.e., dividing the
CCF in 11 slices, but for clear reasons we exclude the continuum level and the level corresponding to the tip of the CCF.}. The values for the upper (near continuum) and
lower bisector points are averaged and subtracted (see Fig.\,\ref{bis13445}). The resulting quantity, the 
Bisector Inverse Slope (BIS), can be used to measure the variations of the line bisector.

\begin{figure}[t]
\psfig{width=\hsize,file=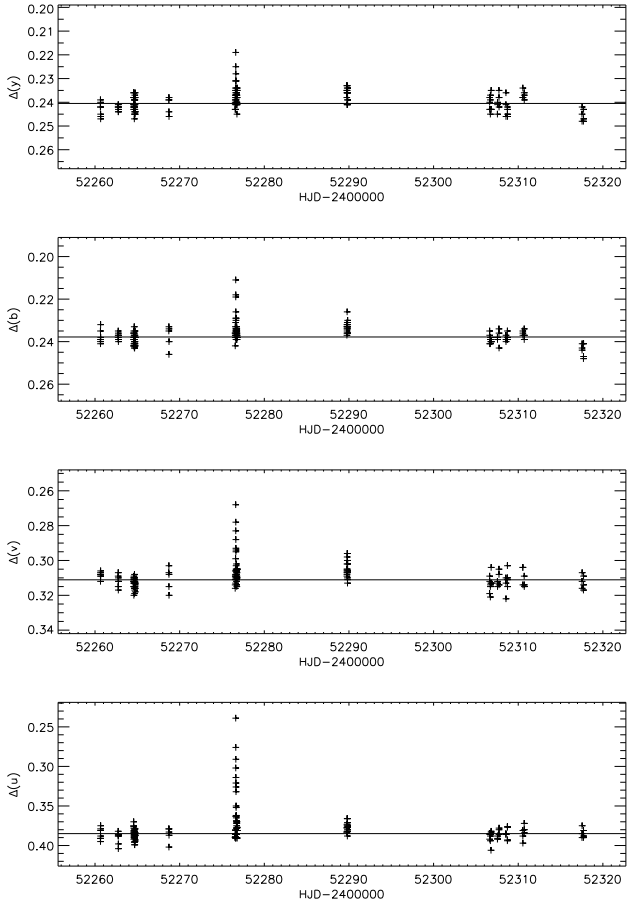}
\caption[]{Magnitude differences \object{HD\,41004\,AB}$-$\object{HD\,39823} in the
instrumental system. The horizontal lines represent the mean values 
(JD\,2452276 excluded), see text}
\label{fig:41004_jd}
\end{figure}

\begin{figure}[t]
\psfig{width=\hsize,file=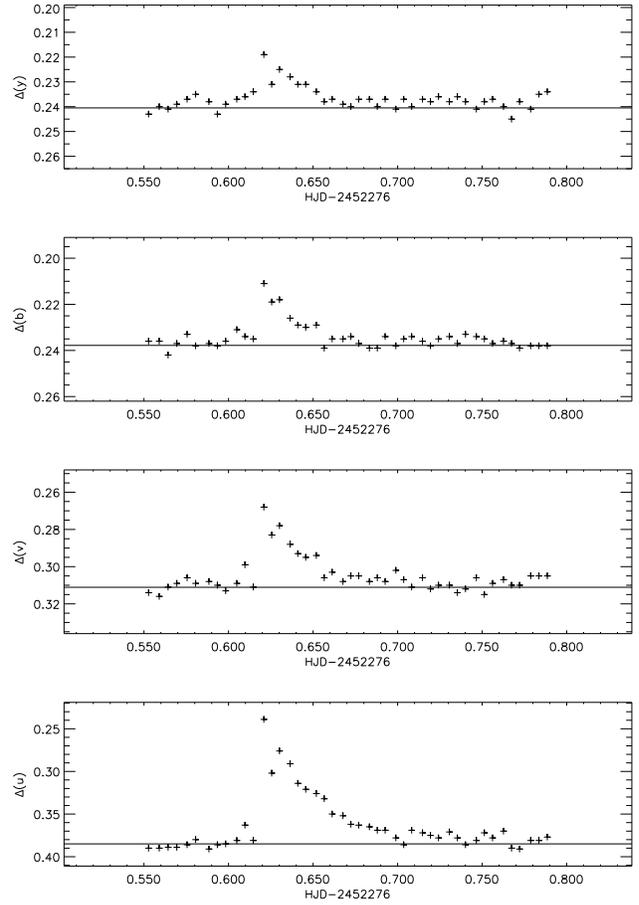}
\caption[]{Magnitude differences \object{HD\,41004\,AB}$-$\object{HD\,39823} in the
instrumental system on JD\,2452276. 
The horizontal lines represent the mean values (JD\,2452276 excluded), see text}
\label{fig:41004_flare}
\end{figure}

The result of the bisector analysis for our star is presented in Fig.\,\ref{fig3}.
As can be seen from the figure, the Bisector Inverse Slope \citep[BIS -- ][]{Que01b} -- 
similar to the usual bisector span \citep{Gra92} -- 
varies in phase with the radial-velocity. This indicates that most probably the radial-velocity
variation observed is being induced by some ``intrinsic'' phenomena, and not by the presence 
of a planet around \object{HD\,41004\,A}. 

We also note that the radial-velocity in this plot was
computed using a different cross-correlation
mask than the one used to determine the radial-velocities presented in 
Fig.\,\ref{fig2} \citep[for more details see][]{Que01b}, and the resulting
radial-velocity variation amplitude is
smaller than the one measured with the standard mask 
(37\,m\,s$^{-1}$ instead of 50\,m\,s$^{-1}$). This large difference would not be 
expected if the radial-velocity signal was induced by the presence of a planet 
around \object{HD\,41004\,A}, since in that case all the lines would ``move'' by the
same amount, leaving the radial-velocity variation (but not necessarily the ``zero'' point)
fairly independent of the mask used. 
The different amplitude observed may be interpreted as a hint that some phenomena is
not affecting in the same way all the spectral lines observed.

\subsection{Photometry}
\label{sec:photom}

The observations described above made us suspect that similarly to \object{HD\,166435},
our star should present photometric variations compatible with the presence of a
rotating spot. According to the Hipparcos catalogue \citep[][]{ESA97}, however, 
\object{HD\,41004\,AB} is stable, with a magnitude scatter
of 0.016\,mag, a value that is typical for a star of its magnitude\footnote{Again, the
photometry is most of all sensitive to the A component of the system, and we are
thus basically measuring the eventual photometric variations of \object{HD\,41004\,A}. For 
example, if the \object{HD\,41004\,B} would change its luminosity by 5\%, the total
flux of the \object{HD\,41004\,AB} system would change by only $\sim$0.15\%, i.e.,
about 0.002 magnitudes, a value that is within the observational errors.}. But this limit is
not very telling, since a spot with a filling factor of a few percent can already induce
a radial-velocity signal with an amplitude of a few m\,s$^{-1}$ in our star \citep[][]{Saa97}.
Furthermore, we could suppose that this star had just developed a spotted region, that was not
present by the time of the Hipparcos measurements.

In order to control the photometric stability\footnote{These observations were also done with the
goal of searching for an eventual planetary transit. 
Given the short period the probability of transit would be particularly high.},  
we have collected precise photometric data for \object{HD\,41004\,AB}. 

%\begin{table}   %empty table, 41004 differential photometry
%\caption[]{\label{tab:41004.lc}
%152 $uvby$ magnitude differences HD41004AB - HD39823
%in the instrumental system.}
%\end{table}

\begin{table}
\caption[]{\label{tab:uvby}Standard $uvby$ indices for HD41004AB and
the comparison stars}
\begin{flushleft}
\begin{tabular}{lrrrrr} \hline \hline
Star     &  $ V$     &   $(b-y)$     &   $m_1$   &   $c_1$   & N \\
\hline
\object{HD41004\,AB}&     8.621 &     0.524     &     0.404 &     0.313 & 49\\
         &$\pm 0.007$&$\pm 0.004$    &$\pm 0.007$&$\pm 0.008$&   \\   
\object{HD38459}  &     8.493 &     0.503     &     0.368 &     0.329 & 30\\
         &$\pm 0.004$&$\pm 0.003$    &$\pm 0.006$&$\pm 0.007$&   \\      
\object{HD39823}  &     8.381 &     0.526     &     0.315 &     0.312 & 38\\
         &$\pm 0.004$&$\pm 0.003$    &$\pm 0.007$&$\pm 0.006$&   \\      
\object{HD43548}  &     8.805 &     0.434     &     0.225 &     0.309 & 26\\
         &$\pm 0.005$&$\pm 0.004$    &$\pm 0.007$&$\pm 0.006$&    \\     
\hline
\end{tabular}
\end{flushleft}
\end{table}

The photometric $uvby$ observations of \object{HD41004AB} were obtained at the Str{\"o}mgren
Automatic Telescope (SAT) at ESO, La Silla, Chile on 11 nights between
December 2001--February 2002. 
Details on the spectrometer and the fully-automatic mode of the telescope 
are given by \citet{Ols93,Ols94}. 

Three comparison stars, \object{HD\,39823} (C1), \object{HD\,38459} (C2), and 
\object{HD\,43548} (C3) were observed together with HD41004AB in the sequence 
C1--\object{HD41004}--C2--\object{HD41004}--C3--\object{HD41004}--C1, 
allowing accurate differential magnitude differences to be formed. 
Each observation consisted of three individual integrations, and the
total number of photo-electrons counted per observation
was at least 75,000 in $u$, and considerably more in the
other three channels. 
Sky measurements were obtained at a position near \object{HD41004\,AB} at least 
once per sequence. 
A circular diaphragm of 17\arcsec\ diameter was used throughout. 
Nightly linear extinction coefficients were determined from the observations 
of the comparison stars and other constant stars, and when appropriate, linear 
or quadratic corrections for drift during the nights were applied.

Differential magnitudes (instrumental system) were formed using for each
candidate observation the two nearest comparison star observations.
All comparison star observations were used with C2 and C3 first shifted 
to the level of C1. A careful check of the comparison stars 
were performed, and during the observing period they were all found to 
be constant within 0.003-0.004 mag. ($vby$) and 0.005-0.006 mag. ($u$), which is
close to the observational accuracy. 

\begin{figure}[t]
\psfig{width=\hsize,file=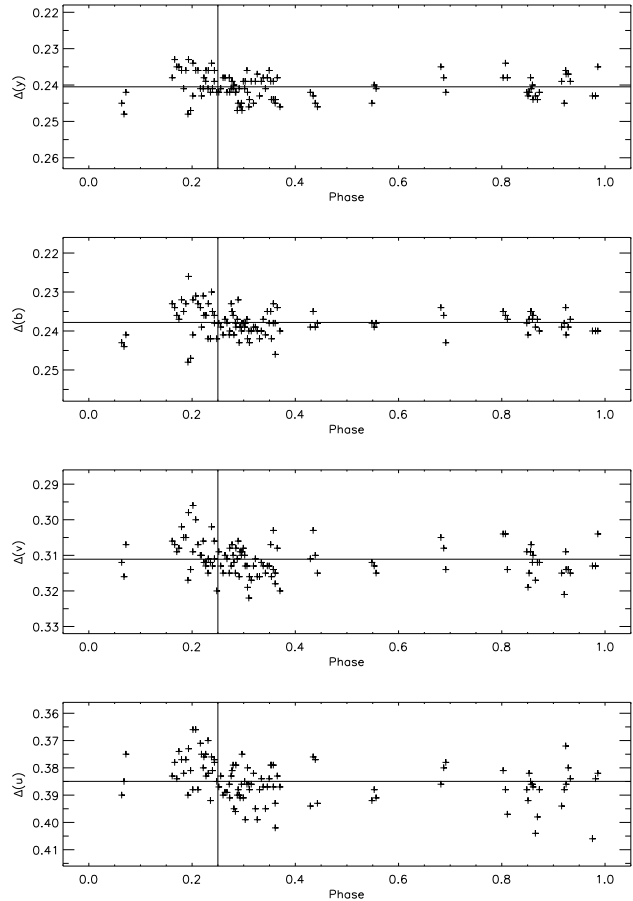}
\caption[]{Phase folded magnitude differences 
\object{HD\,41004\,AB}$-$\object{HD\,39823} in the instrumental
system; JD\,2452276 observations have been excluded.
The horizontal lines represent the mean values, and vertical lines are
drawn at phase 0.25, i.e. at the phase of expected central transit}
\label{fotomphase}
\end{figure}

In Fig.\,\ref{fig:41004_jd} we show the 152 obtained \object{HD\,41004\,AB}$-$\object{HD\,39823} 
magnitude differences in the instrumental system\footnote{The individual 
measurements are available in electronic form at CDS.}.
Typical rms errors of one magnitude difference are
0.004-0.005 ($ybv$) and 0.006-0.008 ($u$).
On one night, JD2452276, a sudden brightening of \object{HD41004\,AB} was observed,
presumably due to a flare. Details are shown in Fig.\,\ref{fig:41004_flare}.

Mean magnitude differences \object{HD\,41004\,AB} - \object{HD\,39823}, 
excluding the JD2452276 observations, are 
$0.240 \pm 0.004$ ($y$), 
$0.238 \pm 0.004$ ($b$), 
$0.311 \pm 0.005$ ($v$), and 
$0.385 \pm 0.008$ ($u$).
Thus, the rms errors are close to those expected from the observational
uncertainties. 
We see no sign of variability correlated with the 1.3 day period (Fig.\,\ref{fotomphase}).
However, as the JD2452276 observations show that HD41004AB is active, long term
variability at some level can not be excluded.

As an extensive list of $uvby$ standard stars were also observed,
an accurate transformation to the standard system could be done. The selection
of standard stars and the transformation to the standard system will
be described in detail elsewhere.
Standard $uvby$ indices are given in Table\,\ref{tab:uvby}.
For \object{HD\,41004\,AB} they are all within 0.035 mag of the mean $uvby$ indices 
for the MK spectral classification K2V.
The colour $(b-y)$ only differs by 0.006 mag from the mean K2V colour
index \citep[cf.][]{Ols84}.

\begin{figure}[t]
\psfig{width=\hsize,file=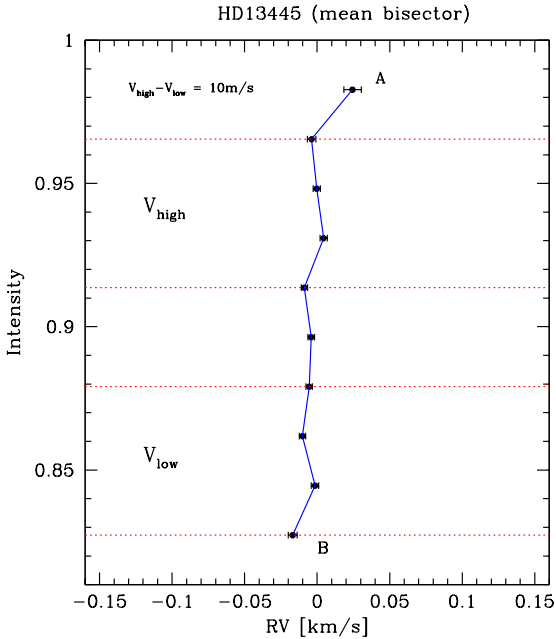}
\caption[]{Average bisector for the K0 dwarf \object{HD\,13445}. This figure 
shows that the line bisector
is almost vertical for a star of this spectral type. The two regions denoted by the
dotted lines (V$_{\mathrm{low}}$ and V$_{\mathrm{high}}$) represent the intervals used
to compute the Bisector Inverse Slope (BIS), defined as BIS $=$ V$_{\mathrm{high}}$$-$V$_{\mathrm{low}}$:
V$_{\mathrm{high/low}}$ are simply the average of the velocity of the 4 ``points'' in each
of the intervals \citep[for more details see][]{Que01b}. The error bars represent the value
of $\sigma(i)/\sqrt(N)$, where $\sigma(i)$ is the rms around the mean velocity 
obtained for a given bisector level $i$ from the $N=118$ CCF's available. A and B 
denote the upper and lower bisector points}
\label{bis13445}
\end{figure}

The results, revealing that the star is stable within the instrumental 
precision\footnote{We note also that no transits were found.}, support the idea that the radial-velocity variation observed cannot be due to the presence of spots in the
surface of \object{HD\,41004\,A}, as was the case for \object{HD\,166435}.
But it is interesting to mention that indeed we would not expect this to be the case, 
given that the rotational period of \object{HD\,41004\,A} (see Table\,\ref{tab1}) is 
much longer than the observed 1.3-day period in radial-velocity\footnote{We note that the
rotational period was obtained from the chromospheric activity level, this latter being 
derived from an analysis of the \ion{Ca}{ii} H and K lines. These lines, located in the
blue part of the spectrum, are probably not affected at all by the
blend with the spectrum of the M dwarf, and will thus give us information
only regarding \object{HD\,41004\,A}.}. Otherwise, and given the low 
$v\,\sin{i}$ for this star, we would be observing a system in a very odd configuration. 
In other words, \object{HD\,41004\,A} is not a
\object{HD\,166435}-analog. This is further supported by the fact that contrarily to 
\object{HD\,166435}, the radial-velocity data varies in phase (and not anti-phase) 
with the BIS (see Fig.\,\ref{fig3}).

\begin{figure}[t]
\psfig{width=\hsize,file=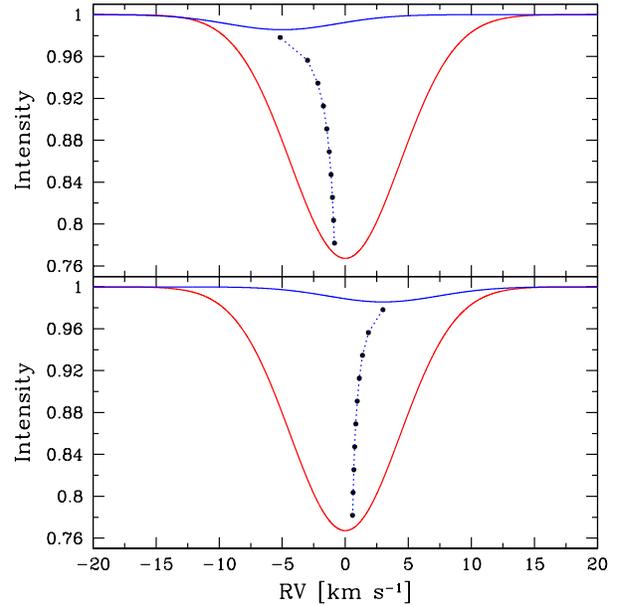}
\caption{Figure illustrating the effect of changing the relative position of the
CCF's of the HD\,41004\,A (deeper CCF) and HD\,41004\,B (smaller CCF): the
resulting blended CCF (the addition of the CCF's of the two stars) will change its profile, with the consequent variations in the observed bisector shape (dashed lines). The CCF's and bisectors in this figure are exaggerated 
for illustration purposes}
\label{shapes}
\end{figure}

\section{Simulating the observations}

The observed correlation between the BIS and the radial-velocity signal, together with
the absence of any significant photometric variation lead us to 
analyze the possibility that the radial-velocity and bisector changes were
being induced by a variation in the relative position of the cross-correlation dips
of \object{HD\,41004\,A} and \object{HD\,41004\,B}. Since 
the angular separation between the two objects is only $\sim$0.5\arcsec,
it is impossible to observe them separately with the 2\arcsec\ fiber
of the CORALIE spectrograph. For each spectrum obtained for \object{HD\,41004\,AB}
we have thus the addition of the spectra arriving from the two components.
The question was thus to know if a periodic change of the relative position of the two 
spectra could induce the observed radial-velocity and bisector variations.
Two main possibilities can be explored in this sense.

\subsection{Case 1}

First, we can consider that
\object{HD\,41004\,A} has a very low mass companion that is responsible for the observed radial-velocity variation, and that \object{HD\,41004\,B} has a ``constant'' velocity.
If that were the case, what would be the effect on the shape of the bisector (i.e. on the BIS)?

A simple simulation, adding two Gaussian functions (i.e. two cross-correlation dips), the first contributing to 
only $\sim$3\% of the 
flux\footnote{This value is based, for simplicity, on the magnitude difference between the two objects.} 
(representing \object{HD\,41004\,B}) and the other with $\sim$97\% of the flux (representing \object{HD\,41004\,A}), 
has shown that changing the position of the deeper function 
by 50\,m\,s$^{-1}$ will reproduce the observed variation in the measured velocity, but leave ``constant'' 
the shape of the bisector. This showed us that we can exclude the possibility that the
observed radial-velocity variation with a period of 1.3\,days is produced by a planetary companion
to HD\,41004\,A.

This simulation suggests that if we hope to simulate the large observed bisector shape variations 
we need to change the relative position of the two cross-correlation dips by (much) more 
than 50\,m\,s$^{-1}$.

\subsection{Case 2}

To try to verify whether large amplitude periodic variation in the relative 
position of the two CCFs can induce the observed signal, we have done a set of simulations where
two cross-correlation functions were added. As above, these two functions, one corresponding to
\object{HD\,41004\,A} and the other to \object{HD\,41004\,B}, were weighted by the 
relative flux of the two stars (A 30 times brighter than B).
These CCF's were considered to be Gaussian for a question of simplicity\footnote{In any case, 
the real shape of the bisector of \object{HD\,41004\,A} is not known.}.
This is a good approximation, since the bisector of a K dwarf is observed to be
quite vertical (Fig.\,\ref{bis13445}) -- see also \citet{Gra92}. However, and as we will see below,
this also poses some problems when trying to constrain the results.

\begin{figure}[t]
\psfig{width=\hsize,file=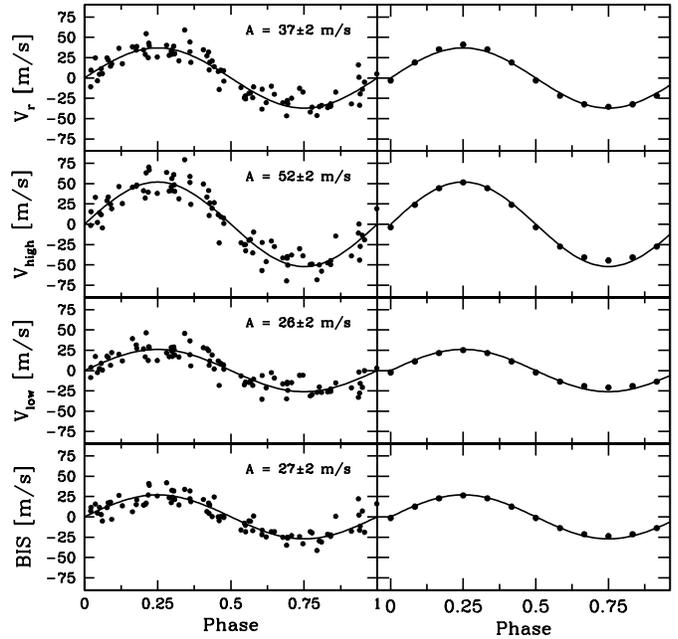}
\caption{Comparison between the observed and predicted amplitudes and variations in 
the radial-velocity, V$_\mathrm{high}$, V$_\mathrm{low}$, and BIS. The best fits to the observed
data are shown (left), as well as the amplitude of the variations ($A$); the fitted curves were 
superposed to the simulated points for comparison. 
For this simulation, the parameters of the secondary CCF were $\gamma_2=-$2.1\,km\,s$^{-1}$, 
$A_2=$0.14, $\sigma_2=$8.0\,km\,s$^{-1}$ and $K_2=$4.0\,km\,s$^{-1}$. The results show the fits 
to be very good}
\label{simul1}
\end{figure}

\begin{figure}[t]
\psfig{width=\hsize,file=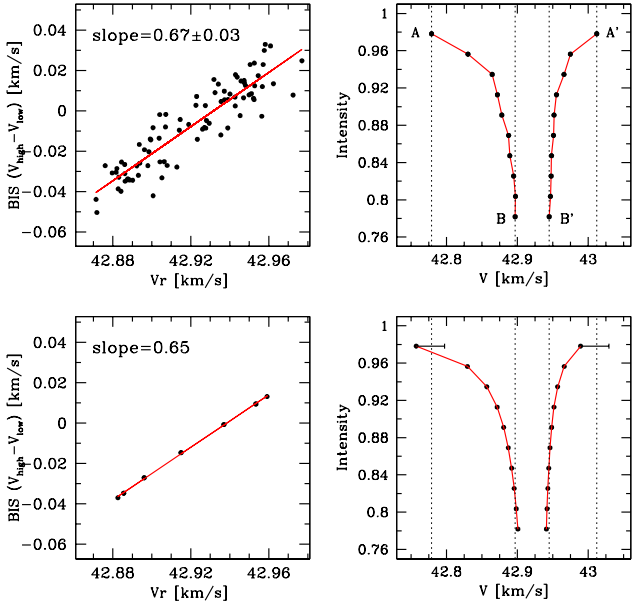}
\caption{{\it Left:} Comparison between the observed and simulated slope of the
relation between radial-velocity and BIS. {\it Right:} Comparison between the
shape of the bisector of the CCF for the maximum and minimum velocities. For the
observed case, an average of the bisectors for the measurements corresponding to the maximum/minimum 
of the velocity was considered. The vertical lines are drawn just to facilitate a comparison.
The error-bars drawn represent the uncertainties in the observed bisector, as well as
the errors due to the fact that the real ``zero'' bisector is not vertical as
considered in the simulations, but slightly ``C'' shaped  (see text for more details).
The parameters of the secondary CCF are the same as in Fig.\,\ref{simul1}. The velocity
scale in these plots is similar to the one of Fig.\,\ref{bis13445}}
\label{simul2}
\end{figure}

In each simulation, the radial-velocity of the primary dip 
(corresponding to \object{HD\,41004\,A}) was considered to be fixed, and the smaller secondary dip,
corresponding to \object{HD\,41004\,B}, was taken to vary periodically with a given
velocity amplitude ($K_2$) (see Fig.\,\ref{shapes}). This variation is supposed to simulate 
the presence of a companion around \object{HD\,41004\,B}. 

A grid of simulations was done by changing both the velocity amplitude $K_2$ (in the range 2-10\,km\,s$^{-1}$) 
and the parameters of the ``small'' dip (the width $\sigma_2$, and the depth $A_2$ were changed in 
the range 5-10\,km\,s$^{-1}$ and 0.04-0.20, respectively), as well as the difference between
the velocity of the primary CCF and the $\gamma$-velocity of the secondary CCF ($\gamma_2$, varied 
between $-$0.5 and $-$3.0\,km\,s$^{-1}$).
In all simulations, the width and depth of the primary dip ($\sigma_1$ and $A_1$) were fixed 
at 4.36\,km\,s$^{-1}$ and 0.24, corresponding to the average parameters of the observed CCF 
of this star (typical parameters for a solar metallicity early K dwarf with low
$v\,\sin{i}$ -- see Appendix).

\subsection{Qualitative results}

A good example of a successful simulation can be seen in 
Figs.\,\ref{simul1} and \ref{simul2}. In the former we compare the observed (left) and simulated (right) 
radial-velocities, V$_\mathrm{high}$, V$_\mathrm{low}$, and BIS for a simulation where the parameters of
the secondary dip are $\gamma_2=-$2.1\,km\,s$^{-1}$, $A_2=$0.14, $\sigma_2=$8.0\,km\,s$^{-1}$ and 
$K_2=$4.0\,km\,s$^{-1}$. The match is extremely good. In Fig.\,\ref{simul2} we compare for the
same simulation the slope of the relation BIS vs. V$_r$ and the actual shape of the bisector
for the maximum and minimum velocities observed and simulated. 

We note that unfortunately the comparison of the bisector shapes (Fig.\,\ref{simul2}, right panels) have a bit 
of an uncertainty, given that we do not know the real shape of the bisector of the CCF of \object{HD\,41004\,A}.
This seems to be particularly true for the bisector point nearest to continuum 
(see Fig.\,\ref{bis13445}). We have thus introduced in Fig.\,\ref{simul2} (lower right panel) 
a tentative error bar in this point, representative of the typical shift observed for the case of
\object{HD\,13445} -- Fig.\,\ref{bis13445}.
If in one hand this particular point is not used 
to compute the BIS, it will change the best general visual bisector shapes.

The important result at this point is that the plots clearly show that with the good set of parameters we can
perfectly fit the observations. But to which extent can we constrain the results in
order to characterize the properties of the CCF of \object{HD\,41004\,B}? In particular,
it would be very interesting to obtain the values of $\sigma_2$ and $K_2$, that would
permit us to derive the projected rotational velocity of \object{HD\,41004\,B}
and the minimum mass for the companion orbiting it.

\subsection{Quantifying the parameters}

The results of the simulations described above have shown that there are some combinations of parameters
for the secondary CCF that can match the observed amplitudes (in radial-velocity, BIS, 
V$_{\mathrm{high}}$ and V$_{\mathrm{low}}$), and the slope of the BIS vs. V$_r$ relation. 
To constrain the models, we have first selected all the simulations
reproducing these quantities within 2-$\sigma$ (corresponding to the uncertainty 
in the observed values, i.e. 4\,m\,s$^{-1}$ for the amplitudes and 0.06 for the slope).

\begin{figure}[t]
\psfig{width=\hsize,file=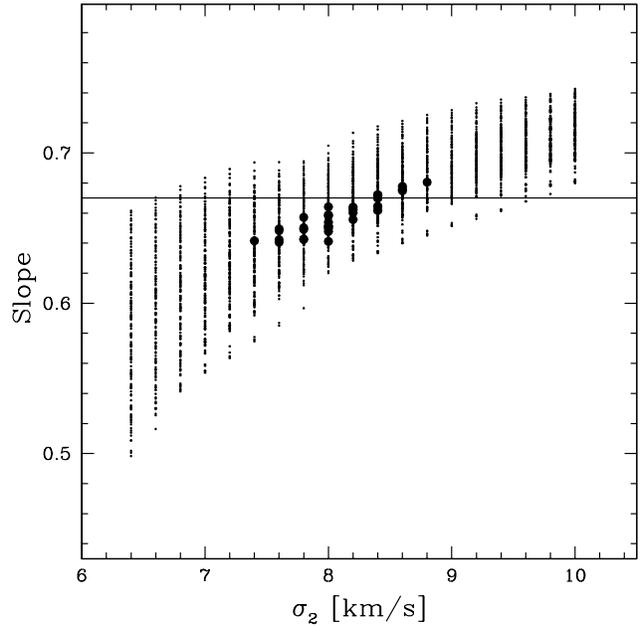}
\caption{Slope of the relation between BIS and $V_r$ against $\sigma_2$ for all the simulations run. We can clearly see the trend of increasing slope with increasing $\sigma_2$. The larger 
dots represent only the simulations whose results are within the 
observational constraints (see text for further details). The line denotes the observed
slope. We clearly see that $\sigma_2$ must have values close to 8\,km\,s$^{-1}$}
\label{a2slo}
\end{figure}

\begin{figure}[t]
\psfig{width=\hsize,file=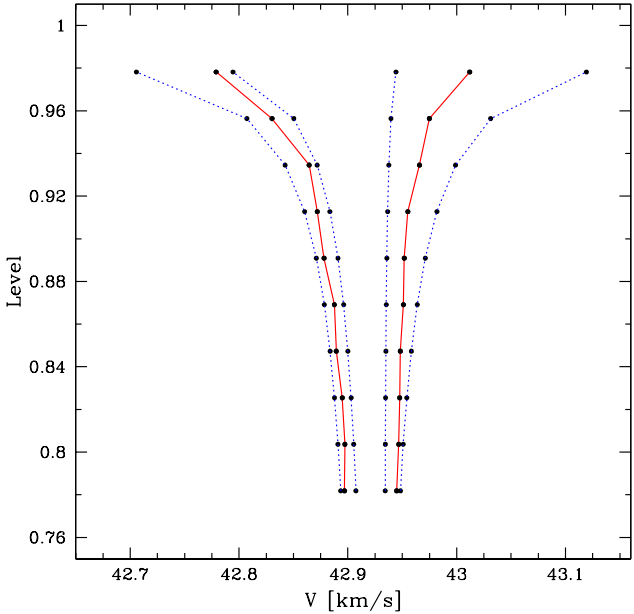}
\caption{Plot of the observed (solid lines) and the simulated (dotted lines) bisectors
illustrating the effect of varying the amplitude $K_2$ in the simulations from 2.5 to 10\,km\,s$^{-1}$. 
All the other parameters are as in Fig.\,\ref{simul1} and \ref{simul2}.}
\label{figshapes}
\end{figure}

We have then further compared the shape and ``stretching'' of the bisectors in order to better constraint the models. For this latter quantity, we have defined
the difference in velocity between the upper bisector point 
(nearest to continuum, lets call it $V(A)$ and $V(A')$ in two bisectors corresponding to extreme velocity cases) 
and lower bisector point ($V(B)$ or $V(B')$) -- see Fig.\,\ref{simul2}. The sum $|V(A)-V(B)|+|V(A')-V(B')|$, 
for example, is quite independent of the initial bisector shape, since it measures a differential 
displacement\footnote{This is true since the bisector 
is probably not strongly ``C'' shaped for a star of this spectral type.}. Based on Fig.\,\ref{bis13445} 
we have considered that we can estimate the observational values of 
$|V(A)-V(B)|$ and $|V(A')-V(B')|$ with a precision of 50\,m\,s$^{-1}$ (to take into account the bisector ``C'' shape), 
and that $|V(A)-V(B)|+|V(A')-V(B')|$ can be obtained with an uncertainty of 30\,m\,s$^{-1}$, a 
value that seems reasonably conservative. But we note again that the absolute shape of the bisectors cannot be used
as a comparison, since the real shape of the bisector for \object{HD\,41004\,A} is not known.

Although there is some degeneracy in the final results, some constraints can be set on the 
parameters of the secondary CCF. 
First, we have verified that the slope of the plot
in Fig.\,\ref{simul2} is most of all dependent on the width of the small dip. 
In all the simulations satisfying the criteria described above,
$\sigma_2$ had values between 7.4\,km\,s$^{-1}$ and 8.6\,km\,s$^{-1}$.
This result, perfectly seen in Fig.\,\ref{a2slo},
permits to determine $\sigma_2$ with a good precision: we 
estimate $\sigma_2$=8$\pm$1\,km\,s$^{-1}$. 
Actually, we can interpret this dependence in light of the fact that the width of the smaller CCF will be the most important 
parameter controlling the extent to which a change in relative position between the two functions
will induce a change in the measured radial-velocity (by fitting a Gaussian function to the
final CCF) and in the BIS$=$V$_\mathrm{high}$$-$V$_\mathrm{low}$.

The knowledge of the width of the CCF of \object{HD\,41004\,B} means that
we can estimate the
projected rotational velocity of this star. Considering that the
CCF of a non-rotating M dwarf with solar-metallicity is $\sim$5\,km\,s$^{-1}$ 
\citep[][]{Del98}\footnote{Such a value
was obtained for the ELODIE spectrograph; with CORALIE, the value
would be a bit smaller, but given the large obtained $\sigma_2$,
the precise knowledge of this value is not crucial.},
using the calibration presented in the Appendix we obtain a value of $\sim$12\,km\,s$^{-1}$ for the
$v\,\sin{i}$ of \object{HD\,41004\,B}. As we will see below, this value will permit to set 
constraints on the mass of the companion to this star (responsible for the 1.3-day period
observed in radial-velocity).

For the other parameters ($K_2$, $A_2$, and $\gamma_2$), the constraints are not so
strong. For example, there seems to be a clear relation that permits to have good models
increasing $A_2$ and decreasing $K_2$.
It would be very interesting, however, to derive $K_2$, since its value could
give us an estimate for the mass of the companion 
orbiting \object{HD\,41004\,B}\footnote{In this sense 
$\gamma_2$ and $A_2$ are not particularly important to constraint.}. First, we have verified 
that it cannot have a very high value, or else we would find 
a flattening in one (or both) extremes of the radial-velocity curve (as already a bit
noticed in the minimum of the the simulated velocities in Fig.\,\ref{simul1}), since the smaller 
CCF would be getting out of the bigger one. Furthermore,
if $K_2$ were very high, the bisector shapes seen in Fig.\,\ref{simul2} would be quite symmetrical (see Fig.\,\ref{figshapes}).  
Based on the simulations satisfying all the criteria described above we can set an upper limit for
$K_2$ around 5.2\,km\,s$^{-1}$ (all the {\it bona-fide} simulations gave $K_2$ between 2.8\,km\,s$^{-1}$ 
and 5.2\,km\,s$^{-1}$). 
From now on we will use the ``worst case'' value of 5.2\,km\,s$^{-1}$, i.e., the value that
will correspond to the higher mass for the companion to \object{HD\,41004\,B}. 
As we will see below, this value will permit to estimate a tentative upper limit for the mass 
of the companion to \object{HD\,41004\,B}.

We could, in principle, try to see the variation in the observed CCF of \object{HD\,41004\,AB}
due to the fact that it is constituted of the sum of two CCF's, one of which changes its position.
To quantify the variation expected, we have subtracted two simulated ``composed'' CCF's
corresponding to the extreme velocity cases of the position of the smaller one. For the parameters 
determined from our simulation, we can show that the difference is of the order of 0.3\% 
(maximum). This small variation, within the observational errors of our measurements, is simply the result of the fact that the CCF of \object{HD\,41004\,B} is very large and shallow.

It is interesting to say that an analysis of the other parameters (width and depth) 
of the observed cross-correlation function of \object{HD\,41004\,AB} has revealed 
no special periodicity nor scatter. 
Such variations could be expected for a typical spectroscopic binary as a consequence of the
variation of the relative position of the CCF's of the two stars. 
In agreement with the observations, our simulations show that these parameters do not change significantly. 
This is basically due to a combination of having $\sigma_2$$>>$$\sigma_1$ and $A_2$$<<$$A_1$.

\section{A brown dwarf around \object{HD\,41004\,B}?}

As shown in the previous section, the best way to explain the
observed radial-velocity signal and CCF-bisector variations is to
consider that \object{HD\,41004\,B} has a companion
in a $\sim$1.3\,day period orbit; considering a stellar mass of 0.4\,M$_{\sun}$ (typical 
for a M2 dwarf), this value corresponds to a separation of only 
0.017\,AU\footnote{I.e., $\sim$7.5 stellar radius, 
considering that \object{HD\,41004\,B} has a radius of 0.5\,R$_{\sun}$, typical for a M2 dwarf.}. 
Taking the maximum amplitude $K_2<5.2$\,km\,s$^{-1}$ deduced from the simulations above, we can compute an upper limit for the minimum mass
of the companion. The results give a value of $\sim$16\,M$_{\mathrm{Jup}}$, strongly 
suggesting that the companion is in the typical brown-dwarf regime.
We note, however, that according to our models, the mass
could even be lower\footnote{Some models that fit the observations have $K_2$ as low as 2.8\,km\,s$^{-1}$,
which would imply a value for the minimum mass of $\sim$8\,M$_{\mathrm{Jup}}$.}, and we cannot completely 
exclude that it is in the
planetary regime. At this moment we prefer to remain cautious on this point,
and we will thus consider this upper limit in the rest of the discussion;
in this case, it is interesting to further discuss the implications of the discovery.

For these calculations we have considered a circular orbit. This is
not only a direct result of the observed and simulated radial-velocity variations,
but it is in fact expected for such a system. Actually, we can estimate the
circularization timescale due to tidal dissipation in the companion's 
convective envelope \citep[][]{Ras96} -- as done for HD\,162020 by \citet[][]{Udr02}. The value
obtained is of the order of 10$^{8}$\,yr, i.e., much shorter than
the stellar age. In other words, we expect the system to be circularized\footnote{We note that 
the circularization timescale due to tidal dissipation in the stellar convective envelope 
\citep[][]{Zah89} is much longer ($\sim$10$^{10}$\,yr).}.

The presence of a third body, in this case \object{HD\,41004\,A}, may induce eccentricity growth
in the system composed of \object{HD\,41004\,B}$+$brown dwarf \citep[][]{Mah79}.
If the period of this eccentricity modulation is shorter than the tidal circularization
period, we could eventually expect that the semi-major axis of the short period system
would decrease, simply because of the tidal dissipation occurring during the
circularization of the orbit. To try to verify if the system could have survived to this effect,
we have computed, using Eq.\,3 in Mazeh \& Shaham, the eccentricity modulation period expected for 
this system. The result gives a value around 1.5$\cdot$10$^{9}$\,yr. This is both longer than the circularization timescale and of the order of the derived stellar age, and thus we can 
expect that the system constituted of \object{HD\,41004\,B}$+$brown dwarf is stable. 
We further note that the timescale for orbital decay due to tidal dissipation 
in the stellar convective envelope is of the order of 10$^{12}$\,yr \citep[][]{Ras96}. 

We can also determine the stellar synchronization timescale for this system (\object{HD\,41004\,B}$+$brown dwarf).
Using the equations in \citet[][]{Udr02}
\footnote{Taken from \citep[][]{Zah89,Zah92}. For our case, 
we have considered the $\lambda_2$, $t_f$, and $k^2$ for a 0.6\,M$_{\sun}$ star, the lower stellar 
mass tabled in
\citet[][]{Zah94}. Since the most important factor determining the synchronization rate is
the ratio between the distance between the two objects and the radius of the star, and that 
these values do not change very steeply with stellar mass, we believe this is a good approximation.}, we have obtained a value of 10$^{7-8}$\,yr. 
It is thus very likely that this
system is synchronized. We can use this fact to estimate the ``true'' rotational velocity
of \object{HD\,41004\,B}, since the synchronization implies that the
rotational period of the star is the same as the orbital period of the companion (1.3\,day).
Taking a stellar radius typical for a M2 dwarf ($\sim$0.5\,R$_{\sun}$), we obtain 
a value of $\sim$20\,km\,s$^{-1}$. 

As we saw above, the width of the
CCF of \object{HD\,41004\,B} can be fairly well constrained by our simulations, permitting to make
a good estimate for the projected rotational velocity $v\,\sin{i}$$\sim$12.0\,km\,s$^{-1}$.
This value implies a $\sin{i}$$\sim$0.6.
Within these conditions, we can deduce that the companion to \object{HD\,41004\,B} 
has a mass lower than $\sim$25\,M$_\mathrm{Jup}$, placing it in the usually considered 
brown-dwarf regime.

At a separation of only $\sim$3.7 Solar radius, we can calculate that the Roche lobe
of the companion is not filled. Using Eq.\,2 in \citet[][]{Egl83}, we obtained a value of
R$_{\mathrm{RL}}$$=$0.18\,$a$, where $a$ is the semi-major axis of the orbit. 
With $a$$=$3.7\,R$_{\sun}$, we obtain R$_{\mathrm{RL}}$$\sim$0.7\,R$_{\sun}$.
This value is a factor of 7 larger than the typical radius of a 25\,M$_{\mathrm{Jup}}$ 
brown dwarf\footnote{At an age of $\sim$1\,Gyr, a
brown dwarf with a mass around 0.03\,M$_{\sun}$ has a radius around 
0.1\,R$_{\sun}$ -- \citet[][]{Cha00}.}. No mass transfer 
is thus expected to occur. 

\section{Concluding remarks}

{
We have presented the case of \object{HD\,41004\,AB}, a system composed of
a K0V star and a 3.7 magnitudes fainter M2 dwarf, separated by 0.5\arcsec. 
Radial-velocity measurements derived from CORALIE blended spectra of the two stars
have unveiled a radial-velocity variation with a period of $\sim$1.3\,days
and a small amplitude ($\sim$50\,m\,s$^{-1}$), compatible with the expected 
signal due to the presence of a planetary companion to \object{HD\,41004\,A}. However, 
as we have seen, the combined radial-velocity, photometry and bisector analysis suggest that
the best explanation for the observations is that the fainter \object{HD\,41004\,B}
has a brown-dwarf companion. In this scenario, the observed low amplitude 
radial-velocity variation is due to the measure of a variation in the line profiles 
which is induced by the change of the relative position of the spectra corresponding 
to the two stellar components \object{HD\,41004\,A} and B.
}

If confirmed, the present discovery represents the first detection of a short-period brown-dwarf 
companion around a M2 dwarf. { In particular, its estimated upper limit mass 
($\sim$25\,M$_{\mathrm{Jup}}$) puts it in the middle of
the so-called brown-dwarf desert, a mass region (between $\sim$20 
and 40\,M$_{\mathrm{Jup}}$) for which (almost) no short period companions to solar-type F, G, 
K \citep[][]{Hal00,Udr00b,Jor01} and M \citep[][]{Mar89} dwarfs were found.} The fact that \object{HD\,41004\,B}
is a M dwarf makes us speculate that the formation of such systems
is more likely for lower mass primaries, i.e. systems having mass ratios
closer to unity \citep[][]{Duq91}. 

\citet[][]{Arm02} proposed a model to explain the existence of the
brown-dwarf desert. One interesting feature of their model is that it predicts that no brown-dwarf
desert should be observed for the companions around the lowest mass dwarfs; they 
set an upper limit of 0.1-0.2\,M$_{\sun}$. Besides the fact that \object{HD\,41004\,B} is slightly 
more massive than this limit, the present case
is interestingly similar to the predictions.

Recently, \citet[][]{Zuc02} discussed an interesting correlation between
the planetary mass and the orbital period. Their analysis strongly suggests that for
single stellar systems, there is a lack of ``high'' mass planetary companions in
short period orbits. On the other hand, Zucker \& Mazeh have found that
for stars in multiple systems, this correlation is no longer present. In particular,
they have pointed out that in these latter cases, there seems to be a negative correlation 
between ``planetary'' mass and orbital period (see their Fig.\,3). It is very interesting 
to see that the companion to \object{HD\,41004\,B} perfectly fits this trend.

A few works have tried to study the formation of planets (or low mass objects) around stars in 
multiple systems. \citet[][]{Nel00} showed that the formation of a planet in a disk is
unlikely for equal mass binary systems with a separation lower than $\sim$50\,AU.
On the other hand, \citet[][]{Bos98} suggested that the influence of a companion
might trigger ``planetary'' formation by disk instability.
Although no strong conclusions are possible at this moment, the fact that \object{HD\,41004\,B} 
is in a double system with a separation that can be as low as $\sim$20\,AU is very interesting 
from the point of view of the formation of its companion. 

Given the uncertainties in the shape of the observed bisector and in the models, 
we do not pretend to have a precise mass determination for the companion to \object{HD\,41004\,B}. Although 
we believe we have obtained a good estimate, one of the main goals of this paper was to illustrate the
importance of combining the radial-velocity data with the bisector analysis. In other words,
one important lesson to be taken from the presented results is that 
the bisector analysis was crucial to correctly interpret the observations.

It is important to caution that this kind of situation can also happen for
long period systems. For those, the signal is very unlikely to have an 
intrinsic stellar activity origin,
but we cannot exclude that it might originate from the presence of a ``wobbling'' stellar companion. 
It is very easy to find a situation where an undetected 
companion, a few tens of an arcsec distant, can be ``contaminating'' our analysis.
Together with a good knowledge of the target star environment, in such a situation the bisector 
analysis seems to provide an unique tool to point out the exact origin of the radial-velocity variations.

In this sense we have analyzed (or re-analyzed) the bisectors for all
the stars with planets discovered in the context of the CORALIE planet search
programme\footnote{See obswww.unige.ch/$\sim$udry/planet/planet.html}. We did not find any 
significant correlation between $V_r$ and BIS. 
We can thus remain confident that the presence of a planetary companion is the best way 
of explaining the radial-velocity variations in these systems. 

\begin{figure}[t]
\psfig{width=\hsize,file=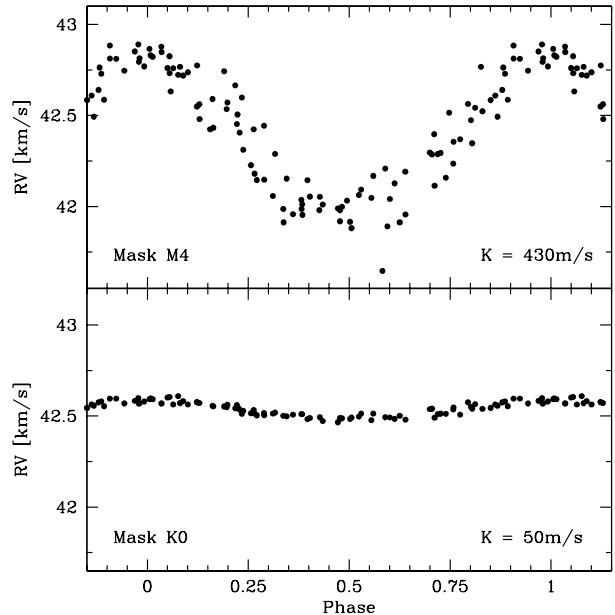}
\caption{Phase folded radial-velocity measurements of \object{HD\,41004\,AB} obtained using 
two different CCF masks. As expected, the amplitudes $K$ obtained vary from mask to mask (see text).
The vertical scales of the two plots is the same to facilitate a comparison.}
\label{fig_masks}
\end{figure}

Besides the bisector analysis, one other way of testing a radial-velocity variation
for such cases may involve the measurement of the radial-velocity using different sets of lines,
or different spectral regions. If the radial-velocity variation is due to the presence of a planet,
we can expect that every spectral region/line will give us about the same velocity amplitude (but not necessarily the same $\gamma$-velocity).
The fact that for \object{HD\,41004\,AB} the amplitude in radial velocity was
different using the cross-correlation mask constructed specially for the bisector analysis \citep[][]{Que01b} 
from the one obtained using the ``classical'' mask is very telling (see Sect.\,\ref{sec:bisector}).
In Fig.\,\ref{fig_masks} we can further see two phase folded diagrams for the radial-velocities
of \object{HD\,41004\,AB} obtained using two different CCF masks. As we can see, if we use a mask specially
constructed for the radial-velocity determination of M4 dwarfs \citep[][]{Del98b}, i.e. a spectral type close to the
one of \object{HD\,41004\,B}, we obtain a much higher amplitude in radial-velocity
than for the case of using a mask constructed for K0 dwarfs. This difference is expected since
in such a case, the difference between the CCF's of \object{HD\,41004\,A} and \object{HD\,41004\,B}
is much smaller (the M dwarf spectrum is enhanced relatively to the K dwarf), being thus the influence 
of the ``small'' CCF stronger. A similar situation (although not as strong) is seen when comparing 
the amplitudes obtained using the K0 and the F0 masks. 

This kind of analysis can, in principle, serve as a test, if the study of the bisector is 
not possible. But we note that for radial-velocity variations induced by the presence of dark spots
in the stellar photosphere, i.e. where the same phenomenon is affecting all spectral lines in
about the same way, we do not expect a strong difference between the radial-velocity amplitudes obtained
using different spectral lines.

Finally, it is interesting to say a few words about other possible ways of confirming the
current detection.
Given the short period of the brown dwarf around \object{HD\,41004\,B}, the probability that
we are able to observe a transit is quite high. In a first approximation, the magnitude variation expected
in such a transit is around 10$^{-3}$ (for the A+B system), a value that does not seem too low. 
But a simpler way of confirming this case would pass by doing high-resolution spectroscopy (and velocity measurements) of \object{HD\,41004\,B}. However, this is not an easy task, since the two components of \object{HD\,41004\,AB}
are separated by only 0.5\arcsec. Unfortunately, there are no available high-resolution spectrographs attached to
an Adaptative Optics system in the southern hemisphere capable of accomplishing this task. The use of 
the Hubble Space Telescope (HST) might represent a solution. Else, the solution may pass
by using high-resolution near-IR spectroscopy, since at those wavelengths the flux of
\object{HD\,41004\,B} is much closer to the one from \object{HD\,41004\,A}.

%
% APPENDIX:
%

\appendix{}
\section{Determination of $v\,\sin{i}$ and [Fe/H] from the Cross-Correlation Function}

One interesting property of the Cross-Correlation Function (CCF) is that any
stellar ``phenomenon'' capable of influencing the lines included in 
the correlation mask (e.g. the line profiles) will be reflected on the CCF itself (i.e. on its
shape, depth and width). In other words, the CCF represents an ``average'' spectral line amidst those 
used in the correlation mask \citep[][]{May85}.
The profile and intensity of the CCF 
is thus a convolution between the profile due to the intrinsic stellar atmospheric 
parameters affecting the lines in the mask (like the global abundance, 
thermal broadening, pressure broadening,
or microturbulence), and the macroscopic macroturbulence 
and rotational velocity of the star \citep{Gra92}, without forgetting the 
instrumental profile, characteristic of a given instrument. 
In the case of CORALIE, given that the correlation mask is constructed using 
mostly weak neutral lines in the spectrum of a standard star (a K0 dwarf), the properties 
of the resulting CCF are basically dependent on the physical quantities controlling the 
properties of a ``typical'' neutral weak metal line. 
For a detailed description of the Cross-Correlation technique we refer to \citet[][]{Bar96} 
and \citet[][]{Pep02}. 

From a practical point of view, this give us the possibility to 
obtain stellar quantities
that are reflected on the spectral lines directly by analyzing the parameters of the
CCF. We can even hope to build 
specific masks for different kinds of lines, each sensitive to a different
stellar parameter. Let us see in the next sections
how this can be used to derive two important quantities from the
CORALIE CCF: the $v\,\sin{i}$ and the stellar metallicity\footnote{For more 
details, and a broader discussion about the errors for the two calibrations presented below,
we refer to \citet[][]{San02}.}.

\subsection{A calibration of the projected rotational velocity $v\,\sin{i}$}
\label{apendix1}

The Gaussian width of a 
weak spectral line of a ``non-rotator''
depends basically on the spectral type and luminosity class.
This is mostly related to the fact that temperature and surface gravity are the 
main variables controlling both the line strength and broadening, due to the temperature and 
pressure effects\footnote{Two other variables, the abundance
and the magnetic field, will be discussed below. Note also that we will ``forget'' about 
natural broadening, since it is constant for each line. Furthermore, we consider that the 
limb-darkening properties are the same for a given temperature and surface gravity.}.
These variables actually indirectly control other broadening parameters 
like the macro- and micro-turbulence \citep{Gra92}.
At the end, the only broadening parameter left that is definitely independent 
of the intrinsic stellar atmosphere properties is the projected rotational velocity 
($v\,\sin{i}$).

Actually, considering that the instrumental profile is constant for each 
star and does not change in time
for a given spectral type and luminosity class (which is a reasonable thing to do for such stable 
instruments like CORALIE) we can expect that every star will have approximately the same 
``zero'' CCF width  -- hereafter called $\sigma_0$ -- and
for a constant temperature and surface gravity, 
the ``excess'' width that the lines (i.e. the CCF) might present is then
mostly due to the effect of the rotational velocity.

\begin{figure}[t]
\psfig{width=\hsize,file=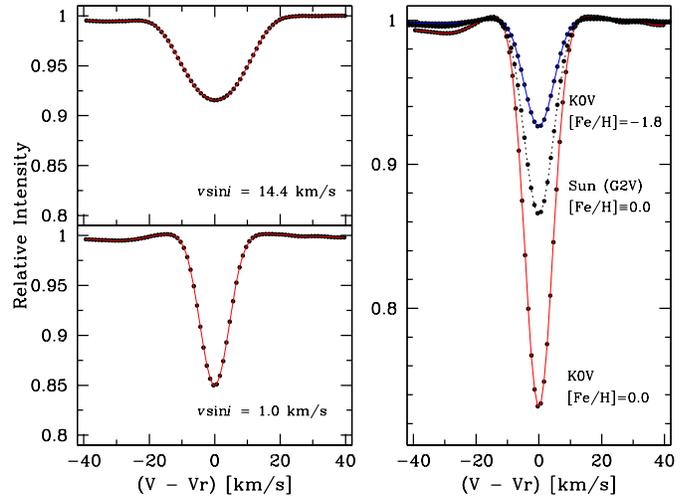}
\caption[]{{\it Left panels}: plot of the CORALIE CCF for two stars with different 
projected rotational velocities. This figure illustrates the effect of the $v\,\sin{i}$ on the CCF;
{\it right panel}: plot illustrating the variation of the surface of the CCF due to changes in
the metallicity and temperature}
\label{ccf_rotation}
\end{figure}

\begin{figure}[t]
\psfig{width=\hsize,file=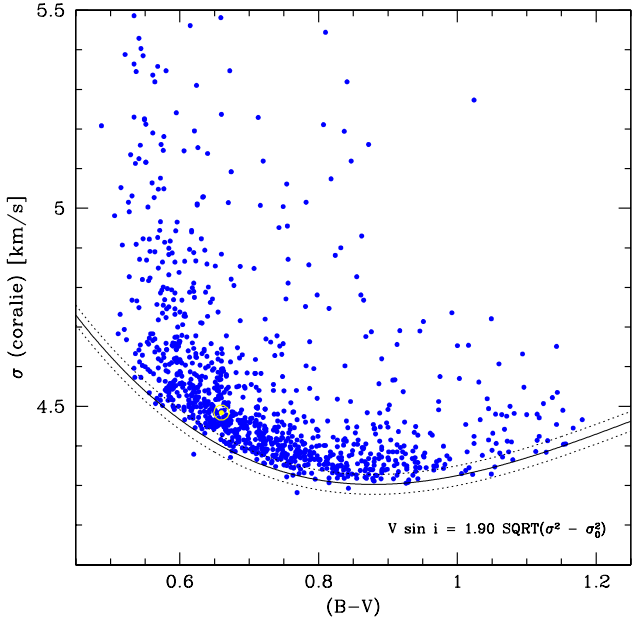}
\caption{Plot of the 
Gaussian width of the CORALIE CCF (expressed in km\,s$^{-1}$) vs. $B-V$. 
The lower envelope (solid line)
determines the locus of the stars with zero rotational velocity.
The two dashed lines represent the uncertainty of 0.025\,km\,s$^{-1}$ in the position of the
lower envelope of the points}
\label{sigma0}
\end{figure}

There are, however, two other physical quantities that can (in principle) change significantly the value of
$\sigma_0$: the metallicity and the magnetic field. In fact, the higher the metal content of
a star, the higher the number of saturated lines; this saturation 
will cause a slight increase in $\sigma_0$. In our case this effect
is not very strong, since the cross-correlation masks are based mainly 
on weak lines. As for the magnetic field, it is known that changing
its intensity will change the Zeeman spliting of the spectral lines, thus
broadening their profiles. This fact can even be used to obtain an estimation of
the stellar magnetic field \citep[e.g.][]{Bor84,Que96}. However, it has been shown 
by \citet{Ben84} that the line broadening due to the effect of the 
magnetic field is only important for late K and M stars. Since this 
effect would be quite complicated to take into account, and it is probably not important for 
most of the stars in the CORALIE survey, we will not consider 
it in the rest of this discussion.

\citet{Ben84} have successfully shown that the
width of the CORAVEL CCF is well correlated with the $v\,\sin{i}$ of 
a star. These two variables can be related by:
\begin{eqnarray}
\label{eqnvsini}
v\,\sin{i} & = & A\,\sqrt{\sigma^2 - \sigma_0^2}
\end{eqnarray}
where $\sigma$ represents the measured Gaussian width of the CCF, $\sigma_0$ the
value of the expected $\sigma$ for a ``non-rotator'' of a given spectral type and luminosity,
and $A$ is a constant relating the ``$\sigma$-excess'' to the
actual projected rotational velocity of the star ($v\,\sin{i}$).
We are, of course, considering that the CCF of a rotating star can
be approximated by a Gaussian. As shown by \citet{Que98} this
is true up to velocities of $\sim$20\,km\,s$^{-1}$, the regime
we are mostly interested in.

Using the same idea, we have used the CORALIE CCF to calibrate a relation
between the $\sigma$ obtained with this instrument and the $v\,\sin{i}$ for solar type dwarfs. 
There are basically two things to do. The first is the determination of $\sigma_0$.
This quantity can be obtained simply by adjusting the lower envelope
of the distribution of points in a plot of $\sigma$ vs. $B-V$ 
(this latter quantity is quite well related with temperature)\footnote{Since the $B-V$ is 
a widespread index available for all the
stars in the CORALIE sample, it is a particularly suitable variable to use.}. 
In this envelope we will, in principle, find the non-rotator stars.
We note that most of the stars observed with CORALIE are dwarfs, 
and there are probably no objects with luminosity classes lower than IV. 
Such a plot is seen in Fig.\,\ref{sigma0}, and the adjusted lower envelope 
is well described by:
\begin{eqnarray}
\label{eqnsigma0}
\sigma_0 & = & 6.603-6.357\,(B-V)\\
& & +5.533\,(B-V)^2-1.454\,(B-V)^3 \nonumber
\end{eqnarray}

It is interesting to say a few words about Eqn.\,\ref{eqnsigma0}. 
As can be seen from Fig.\,\ref{sigma0}, the fitted function decreases up to values of 
$B-V$$\sim$0.9 (where it has a minimum) and then increases again towards higher $B-V$ values. 
In the low $B-V$ regime, this decrease might simply be explained by a decrease in the
macroturbulent dispersion \citep[][]{Gra92}.
On the other side, for high $B-V$ values, the increase in $\sigma_0$ 
is probably due to the increase in the damping constant due to van der Waals
broadening, that can be shown to increase with decreasing temperature. Furthermore, up 
to a spectral type of K5, 
neutral metallic lines increase in strength, thus increasing the number
of saturated lines (for a constant abundance). 
As discussed in \citet[][]{Ben84}, the increase of the
magnetic field as a funcion of increasing spectral type might also
be responsible for the increase in $\sigma_0$ after a given spectral type. 
Together with these physical processes, variations resulting from the way the CCF mask is optimized 
\citep[][]{Bar79,Ben84} might also
introduce some trend\footnote{We expect $\sigma$ to be slightly different in the red and in the blue
part of the spectrum since the ``lines'' in the mask are in average slightly 
larger in the red; consequently, this might introduce a small increase
of the CCF with increasing spectral type.}. It is, however, difficult to
quantify exactly what is the net effect of all these processes.

It is important to mention that after all the metallicity does not seem to play an important 
role in the determination of $\sigma_0$. A few tests were done, separating
the stars in Fig.\,\ref{sigma0} into different metallicity bins. 
The result showed that the effect is very small, and difficult to 
quantify amidst the noise in the lower envelope of the points. 
The interpretation of this is, as mentioned above, probably 
quite simple: since we are dealing with
weak lines in the linear part of the curve of growth, the saturation effect is
not expected to be very strong.

The other variable in equation\,\ref{eqnvsini} we need to determine is $A$, i.e., 
the constant relating the $v\,\sin{i}$ to the ``excess'' width of the CCF. This can 
be done in the same way as in \citet{Que98}: by measuring the 
variation of $\sigma$ as we convolve the CCF of non-rotating stars
with a given rotational profile \citep{Gra92}. For CORALIE we find 
$A$$=$1.9$\pm$0.1, close to the value obtained for ELODIE \citep{Que98}.
In any case, the final value of $v\,\sin{i}$ is not strongly dependent on
errors in $A$.

The technique described above represents a very simple way of obtaining precise
projected rotational velocities for dwarfs, simply as a by-product of
the precise radial-velocity measurements. In particular, the calibration 
permits us to obtain quite easily values for the $v\,\sin{i}$ for all the stars in 
the CORALIE planet search programme.

\begin{figure}[t]
\psfig{width=\hsize,file=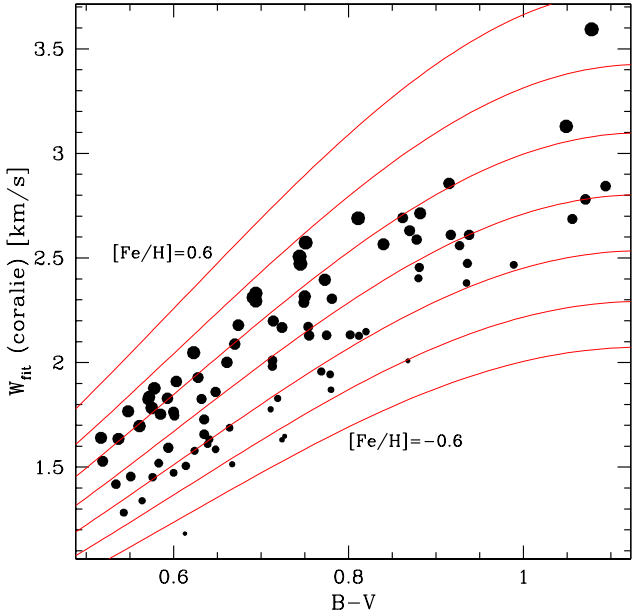}
\caption{Plot of the CORALIE CCF surface 
$W_{fit}$ as a function of $B-V$. $W_{fit}$, expressed in km\,s$^{-1}$, is defined
as $\sqrt{2\,\pi}\cdot\sigma_{fit}\cdot\,A_{fit}$, where $\sigma_{fit}$ and $A_{fit}$ 
are the Gaussian width (in km\,s$^{-1}$) and depth of the measured CCF, respectively. 
The lines represent regions with [Fe/H]=constant. 
The size of the points is proportional to the spectroscopically determined metallicity.
The spectroscopic metallicities were taken from \citet{San01,San01b}}
\label{metalcalib}
\end{figure}

\begin{figure}[t]
\psfig{width=\hsize,file=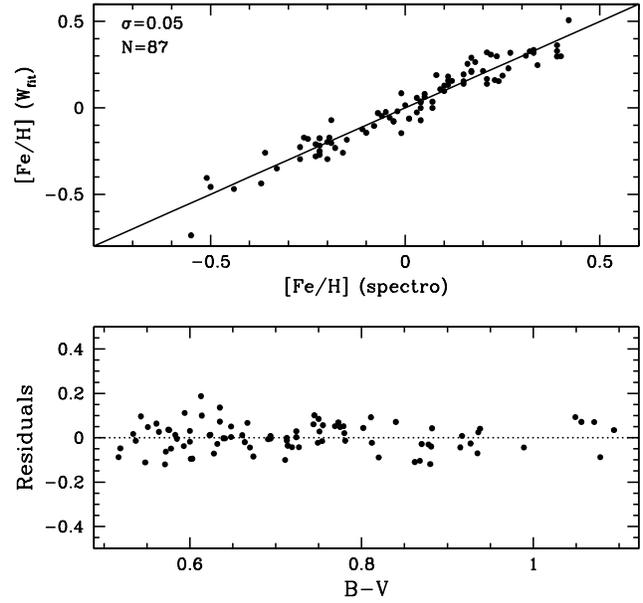}
\caption[]{{\it Upper panel}: Comparison between the spectroscopic and calibrated [Fe/H]. 
{\it Lower panel}: Residuals of the calibration as a function of $B-V$; no dependence is found. The spectroscopic metallicities were taken from \citet{San01,San01b}}
\label{rmsmetalcalib}
\end{figure}

\subsection{A calibration of the metallicity}
\label{apendix2}

Another variable that can easily be obtained from the CCF of a star is
its metallicity. This has been firstly 
done by \citet{May80} for the CORAVEL CCF, a calibration that was later 
improved by \citet{Pon97}.

As mentioned above, the mask used for CORALIE is mainly built out of
a set of weak neutral spectral lines (from a template K0 spectra).
Since most of them are iron lines (the main contributor for the
line opacity in solar type stars), we can expect that
the surface (i.e. the Equivalent Width) of the CCF is well
related to the [Fe/H] of a star.

Furthermore, the fact that we are dealing with weak neutral lines in solar-type stars
implies that the surface of the CCF will be basically independent
on the microturbulence and the surface gravity. These would have an
important effect if we were not dealing with lines that are
in the linear part of the curve of growth, since the broadening
produced by these effects would act to de-saturate the lines, increasing thus
their equivalent widths. Furthermore, it can be shown that
weak lines of an element for which most of it is in the next ionization 
state (most of the iron in a solar-type star is in the form of \ion{Fe}{ii})
are quite insensitive to pressure changes. The surface of the CCF will 
thus mainly depend on the 
temperature (well correlated with $B-V$) and abundance (see Fig\,\ref{ccf_rotation}).

To calibrate a relation between the surface of the CORALIE CCF (hereafter $W_{\mathrm{fit}}$) 
and the iron abundance expressed by [Fe/H], we have used a set of stars for 
which we had obtained
precise [Fe/H] values using a detailed spectroscopic analysis \citep{San01,San01b}. 
The fit, shown in Fig.\,\ref{metalcalib} gives:
\begin{eqnarray}
\label{fehcor}
%[Fe/H] & = & \frac{\log{W_{cor}}+0.602-1.879\,(B-V)+0.846\,(B-V)^2}{0.214}
%[Fe/H] & = & 2.813+4.673\,\log{W_{cor}}\\
%       &   & -8.780\,(B-V)+3.953\,(B-V)^2 \nonumber
[Fe/H] & = & 2.573+4.587\,\log{W_{cor}}\\
       &   & -8.142\,(B-V)+3.583\,(B-V)^2 \nonumber\end{eqnarray}
a calibration valid for $0.52<B-V<1.1$, $1.18<W_{\mathrm{fit}}<3.59$ and $-0.55<[Fe/H]<0.42$.
As expected, the figure shows that for a given metallicity, the equivalent width of the spectral lines (i.e. the surface of the CCF) increases with decreasing temperature. This is the ``normal'' behaviour of a weak metallic line in this temperature regime (up to spectral types around K5, $B-V$$\sim$1.2).

A comparison between the metallicities obtained from the CCF surface and a spectroscopic analysis
shows that the fit has 
a remarkable small dispersion (0.05\,dex -- Fig.\,\ref{rmsmetalcalib}), similar to the uncertainties 
in the spectroscopic determinations. Furthermore, there seems to be no special
systematics with temperature.
We can thus trust on this calibration to obtain precise values of [Fe/H] for
the stars in the CORALIE planet search programme without any long and arduous 
spectroscopic analysis. We note that the metallicities obtained this way are
spectroscopic determinations, since we are using spectral line information. 
Although we cannot exclude that for a particular star the result might
not be as precise as a detailed spectroscopic analysis determination, at least in 
statistical studies this is certainly an extremely accurate technique.

\begin{acknowledgements}
  We want to thank T. Mazeh for the interesting discussions and suggestion, and
  S. Bouley and R. Leguet for having kindly observed HD\,41004\,AB at the SAT on
  two nights each, as well as the anonymous referee whose comments helped
  a lot to improve the clarity of the paper. We wish to thank the Swiss National 
  Science Foundation (Swiss NSF) for the continuous support for this project. 
  The photometry was obtained as part of an extensive study of GK type
  eclipsing binaries supported by the Danish Natural Science Research
  Council. Support from Funda\c{c}\~ao para a Ci\^encia e Tecnologia, Portugal, 
  to N.C.S. in the form of a scholarship is gratefully acknowledged.
  This research has made use of the Simbad database, operated
  at CDS, Strasbourg, France.
\end{acknowledgements}

\end{document}